\documentclass[preprint,aps]{revtex4}
\usepackage{graphicx}
\usepackage{amsfonts}
\begin{document}
\title{$U(1)$ Problem Revisited}
\author{H.Banerjee}
\author{Gautam Bhattacharya}
\email{gautam@theory.saha.ernet.in}
\affiliation{Saha Institute of Nuclear Physics\\
1/AF Bidhannagar, Kolkata 700064, INDIA}

\begin{abstract}
In the anomaly equation for the singlet axial current the chiral limit of the quark mass term does not vanish but comprises contribution from  fermion zero  modes whose integral exactly cancels the topological charge arising from the Adler-Bell-Jackiw anomaly. This signals chiral symmetry and opens a window for restoring the status of Goldstone boson for the singlet $\eta^\prime$ without having to invoke the large $N_c$ limit in the underlying QCD. We construct the anomaly term in the effective action that incorporates the chiral symmetry property  and yet accounts for the excess mass of $\eta^\prime$ only when chiral symmetry is broken explicitly by the quark masses. The anomaly term in the present scenario thus plays the role of a catalytic agent that enhances the mass of $\eta^\prime$ so that the singlet axial current obeys the popular PCAC condition.
\end{abstract}
\pacs{11.10Ef,11.30Fs,11.30Rd,12.39Fe}
\maketitle
\section{Introduction}
In the popular scenario \cite{gellmann} of spontaneous breaking of chiral symmetry
through  a nonzero quark condensate in QCD vacuum the lightest
pseudoscalar mesons $\pi, K$ and $\eta$ are the natural candidates for the
Goldstone Bosons associated with the octet of quark currents with
flavours. Massless in the chiral limit, they acquire their observed
masses as the light quarks $u$, $d$ and $s$ are given small masses in the
underlying QCD Lagrangian. It is a moot question if a singlet $\eta^\prime$
as heavy as 1 Gev should be identified as the Goldstone Boson for the singlet 
axial current $\bar q \gamma_\mu\gamma_5 q$. More challenging is the question whether 
the Adler-Bell-Jackiw (ABJ)  anomaly in the singlet current is an aberration that does not really destroy
chiral symmetry, i.e. invariance under global chiral transformation, and
hence the status of $\eta^\prime$ as the singlet Goldstone Boson but only
fulfill the role of  the missing element needed to explain its excess mass.
All these issues constitute what is known as the $U(1)$ problem.

A turning point in the long tortuous history \cite{christos} of $U(1)$ 
problem is 't Hooft's \cite{thooft} observation that, though a total divergence, the ABJ
anomaly can have nonzero space-time integral arising from instanton configurations
of the gluon field and thus may induce large physical effects like the mass of 
$\eta^\prime$. This essentially constituted the basis of the perception \cite{witten,veneziano1,veneziano} that chiral limit alone can not 
restore chiral symmetry
and the status  of Goldstone boson for $\eta^\prime$. One should, in
addition, take recourse to the large $N_c$ (number of colours) limit in
which quark loops necessary to generate ABJ anomaly are suppressed.

The present revisit is inspired by the recognition that ABJ anomaly
representing the topological charge density does not saturate the chiral
limit of the divergence of the singlet axial current in an instanton
background. The pseudoscalar mass term $(m\bar q\gamma_5 q)$ has nontrivial
chiral limit arising from zero modes spawned in the fermion sector by
instantons. These extra contributions are precisely what one needs to
cancel, thanks to the Atiah-Singer index theorem, the topological charge
contributed by ABJ anomaly to the chiral limit of the integral of the
divergence of the singlet axial current. By ensuring invariance under
`rigid'(space-time independent) chiral transformations in underlying QCD, 
this result opens a window for the restoration of the status of Goldstone
Boson for singlet $\eta^\prime$ in the chiral limit without having to invoke
the large $N_c$ limit. Would it then be possible to interpret and understand  
the excess mass of $\eta^\prime$ through suitable terms \cite{derujula} 
in the effective Lagrangian to represent the anomalous contributions of
triangle diagram in the underlying QCD? We wish to address these issues in
the present revisit.

\section{Chiral limit}
For clarity of our discussion it is convenient to introduce Pauli-Villars
fermions $\chi$ of mass $M$  to regularise quark loops. In path integral
framework the Jacobian for chiral transformation is trivial 
\cite{fujikawa,banerjee1} with this regularisation and the 
anomalous Ward identity for the singlet axial current of quarks of $L$ flavours 
each with mass $m$ in the underlying QCD assumes the form
\begin{equation}
<\partial_\mu J_{\mu5}>=2L(D(x)-Q(x)),\label{anomaly}
\end{equation}
\noindent with
\begin{equation}
\begin{array}{l}
\displaystyle{D(x)=m<\bar q(x)  \gamma_5 q(x)>,}\\
\displaystyle{Q(x)=\lim_{M\rightarrow\infty}M<\bar\chi(x)\gamma_5\chi(x)>.}\label{DQ} 
\end{array}
\end{equation}
Fermion averaging  $< >$ in (\ref{DQ}) is to be implemented in the
orthonormal eigenbasis ${\phi_n}$ of the hermitian Dirac operator $D\!\!\!\!/$
\begin{equation}
\begin{array}{ll}
\displaystyle{D\!\!\!\!/\equiv \gamma^\mu(i\partial_\mu - gA_\mu),}&\displaystyle{A_\mu=A^a_\mu t^a;}\\
\displaystyle{D\!\!\!\!/\phi_n=\lambda_n\phi_n,}&\displaystyle{\int d^4x\phi_m^\dagger(x)\phi_n(x)=\delta_{mn}}
\end{array}
\label{definition}
\end{equation}
One obtains \cite{banerjee2}
\begin{equation}
\begin{array}{l}
\displaystyle{D(x)=m\sum_n\frac{\phi_n^\dagger(x)\gamma_5\phi_n(x)}{m+i\lambda_n},}\\ 
\\
\displaystyle{Q(x)=\lim_{M\rightarrow\infty}M\sum_n\frac{\phi_n^\dagger(x)\gamma_5\phi_n(x)}{M+i\lambda_n}}\label{PV}
\end{array}
\end{equation}
Nonzero eigenvalues $\lambda_n$ have chiral partners $-\lambda_n$ belonging 
to the eigenmode $\gamma_5\phi_n$. The zero eigenmodes
\begin{equation}
D\!\!\!\!/\phi_{0i}=0,\ \ \gamma_5\phi_{0i}=\epsilon_i\phi_{0i},\label{zerom}
\end{equation}
however, have definite chirality with $\epsilon_i=\pm 1$.

Despite appearances, the chiral limit of the pseudoscalar mass term $D(x)$
does not vanish and is instead composed of the zero modes

\begin{equation}
\lim_{m\rightarrow 0}D(x)=\sum\epsilon_i\phi^\dagger_{0i}\phi_{0i}
\end{equation}
On the other hand, in the asymptotic $M\rightarrow\infty$ limit of the Pauli-Villars
term one recognises the exponent of the Jacobian obtained by Fujikawa \cite{fujikawa} 
for chiral transformation
\begin{equation}
\begin{array}{ll}
Q(x)&=\displaystyle{\sum_n\phi_n^\dagger(x)\gamma_5\phi_n(x)}\\
&=\displaystyle{\frac{g^2}{32\pi^2}F_{\mu\nu}\tilde F_{\mu\nu}}\label{adler}
\end{array}
\end{equation}
One is thus persuaded to rewrite Eq.(\ref{anomaly}) as
\begin{equation}
<\partial_\mu J_{\mu5}>=2L(D_m(x)-A(x))
\end{equation}\label{anomaly2}
with
\begin{equation}
\begin{array}{ll}
D_m(x)&=\displaystyle{D(x)-\sum\epsilon_i\phi^\dagger_{0i}\phi_{0i},}\\
A(x)&=\displaystyle{Q(x)-\sum\epsilon_i\phi^\dagger_{0i}\phi_{0i}}\\
&=\displaystyle{\frac{g^2}{32\pi^2}F_{\mu\nu}\tilde F_{\mu\nu}-\sum\epsilon_i\phi^\dagger_{0i}\phi_{0i}}
\end{array}\label{chiraldef}
\end{equation}
to make the chiral limit transparent
\begin{equation}
<\partial_\mu J_{\mu5}>_{m=0}=-2LA(x)\label{anomaly3}
\end{equation}
Thus the total anomaly $A(x)$ comprises two pieces \cite{jackiw} of which the first is the 
familiar ABJ term Eq.(\ref{adler}) and the other is the contributions from zero modes induced
in the fermion sector by instantons. Thanks to the Atiyah-Singer index theorem
\begin{equation}
\nu=n_+-n_- \label{index}
\end{equation}
with winding number $\nu$  given by 
\begin{equation}
\nu=\frac{g^2}{32\pi^2}\int {d^4x F_{\mu\nu}\tilde F_{\mu\nu}}\label{winding}
\end{equation}
and $n_+(n_-)$ the number zero modes of positive(negative) chirality, the total
integral of the divergence of the singlet axial current vanishes 
\begin{equation}
\int{d^4 x <\partial_\mu J_{\mu_5}>_{m=0}}=0
\label{conservation}
\end{equation}
in the chiral limit.

Eq.(\ref{conservation}) signals chiral symmetry in the underlying QCD which played earlier \cite{banerjee1}
a key role in demonstrating that in QCD a $U(1)$ chiral phase in quark mass matrix is 
unphysical. This led to a resolution of the strong CP problem.
The unphysicality of the $U(1)$ chiral phase is obvious and transparent in the Minkowski metric. A similarity transformation of the Dirac matrices
\begin{equation}
\gamma_\mu \rightarrow \gamma_\mu^\prime= e^{-i\gamma_5\frac{\theta}{2}}\gamma_\mu e^{i\gamma_5\frac{\theta}{2}}
\end{equation}
eliminates the $U(1)$ chiral phase $\theta$ in quark masses in QCD Lagrangian.

It is important to recognise that the anomaly $A(x)$ is nonlocal. This is not surprising. Anomaly equations are not operator identities but are results of regularisation that emerge only after the regulators are removed. For the singlet axial anomaly $A(x)$ the representation in Eq.(\ref{chiraldef}) appears only when the ultraviolet regulator (the Pauli-Villars mass $M$) and the infrared regulator (the quark mass $m$) approach their appropriate limits in the right hand side of Eq.(\ref{anomaly})
\begin{equation}
A(x)=-<m\bar q(x)  \gamma_5 q(x)-M\bar \chi(x)  \gamma_5 \chi(x)>_{m=0,M=\infty}\label{rightdef}
\end{equation}
and fermion averaging $< \ >$ is implemented in  path integral framework. In the light of these observations one would believe that grafting anomaly equations into operator Ward identities may not be quite warranted and may have sowed the seeds of controversies \cite{thooft1} surrounding the $U(1)$ problem.

One obtains in path integral framework the correlator equation
\begin{equation}
\left<(\int d^4x\partial_\mu J_{\mu 5}(x))(\bar q \gamma_5 q(0))\right>_{m=0}=-2L\left<(\int d^4xA(x))(\bar q \gamma_5 q(0))\right> +<\bar q q> = <\bar q q> \label{ward}
\end{equation}
which follows from chiral symmetry, Eq.(\ref{conservation}), of the QCD action. This is the analog of the Ward identity discussed in literature \cite{christos,thooft1} in the context of realisability of the singlet $\eta^\prime$ as a Goldstone boson in the chiral limit. Eq.(\ref{ward}) suggests the presence of a massless pseudoscalar particle if the quark condensate term $<\bar q q>$ is nonvanishing. This opens the window for restoring the status of Goldstone boson for the singlet $\eta^\prime$.

The stage is thus set for addressing the problem of realising the excess mass of $\eta^\prime$ as and when one moves away from chiral limit by giving small masses to quarks. It is but natural to attribute the excess mass to contributions from a term in the effective Lagrangian that represents the anomaly in the singlet axial current and yet obeys the constraint, Eq.(\ref{conservation}), in the chiral limit.

\vspace{24pt}
\section{Anomaly Terms in the Effective Action}

According to  Eq.(\ref{conservation}) the global(rigid) U(1) chiral symmetry is unaltered even when the divergence of axial vector current is anomalous. Therefore, it is perfectly natural to regard the singlet pseudoscalar meson $\eta^\prime$ as a Goldstone boson in the chiral limit (i.e. when the quarks are massless) just like its flavour counterparts. One can then include this `ninth' ($L^2$th to be precise) component in the chiral model -- the effective low energy manifestation of QCD involving only the pseudoscalar fields and their currents. Here the spontaneously broken chiral symmetry is realised through a nonlinear sigma model of matrix-valued fields associated with the flavour group $SU(L)$. The leading term of the action 
\begin{equation}
\begin{array}{l}
S_0=-F_\pi^2 \int d^4x {\rm Tr}(\partial_\mu {\cal M}\partial^\mu{\cal M})=-F_\pi^2 \int d^4x {\cal L}_0,  \ \ {\cal M}\in U(L)\\
{\cal M}=\displaystyle{\exp(i\frac{\eta^\prime}{F_\pi}){\cal N}, \ \ {\cal N}\in SU(L)}.
\end{array}
\label{seed}
\end{equation}
describes the the dynamics of $U_{\rm Left}(L)\times U_{\rm Right}(L)$ symmetry broken spontaneously to $U_V(L)$ along with the conservation of $L^2$ axial vector currents ${\cal M}^{-1}\partial_\mu {\cal M}$. 

The action in Eq.(\ref{seed}) is consistent with the global symmetries of massless QCD including the $U(1)$ axial symmetry, but does not reflect the anomaly relation as given in Eq.(\ref{anomaly3}) for the colour gauge invariant singlet axial current. This lacuna can be cured by introducing the anomaly term in the Lagrangian 
\begin{equation}
{\cal L}_1=\frac{L}{F_\pi}\partial_\mu \eta^\prime K^\mu\label{anomalyterm}
\end{equation}
where for $K_\mu$ the natural ansatz \cite{witten,veneziano1,veneziano} is
\begin{equation}
K^\mu=\frac{g^2}{16\pi^2}\epsilon^{\mu\nu\rho\sigma}(A^a_\nu\partial_\rho A^a_\sigma +\frac{g}{3}f^{abc}A^a_\nu 
A^b_\rho A^c_\sigma)\label{kmudef}
\end{equation}
whose four divergence is the familiar ABJ term
\begin{equation}
\partial_\mu K_\mu=\displaystyle{\frac{g^2}{16\pi^2}F_{\mu\nu}\tilde F_{\mu\nu}}
\end{equation}

At first sight it may appear that the inclusion of Eq.(\ref{anomalyterm}) in the effective Lagrangian would yield only  the ABJ term in the expression for the divergence of the singlet axial current contradicting Eq.(\ref{conservation}).  This apparent discrepancy melts away if one recognises the nontrivial contribution from surface in the  variation of the action under $\eta^\prime(x)\rightarrow\eta^\prime(x)+\theta(x)$
\begin{equation}
\delta S_1 = \int d^4x\frac{\partial {\cal L}_1}{\partial(\partial_\mu\eta^\prime)}(\partial_\mu\theta)\\
=\int d^4x\partial_\mu(\theta K_\mu)\\
- \int d^4x
\theta(x)\partial_\mu\frac{\partial {\cal L}_1}{\partial(\partial_\mu\eta^\prime)}\label{variation}
\end{equation}

It should be observed that the representation Eq.(\ref{anomalyterm}) for the anomaly term enjoys translation invariance $\eta^\prime\rightarrow \eta^\prime+\theta$, which, on the one hand reflects the chiral symmetry of the undelying QCD and on the other hand ensures  that $\eta^\prime$ is massless in the chiral limit. In contrast, the alternative choice $\eta^\prime\partial_\mu K_\mu $ popular in literature \cite{witten, veneziano,veneziano1} reflects loss of both chiral symmetry and masslessness of $\eta^\prime$. The contributions from the fermion zero modes \cite{jackiw} which play the key role in realising chiral symmetry even in the presence of ABJ anomaly seem to have been ignored in literature \footnote{In the review\cite{christos}  both the choices are mentioned but without any reference to the fermion zero modes and chiral symmetry in the underlying QCD}.  Clearly the two terms $\eta^\prime\partial_\mu K_\mu $ and $K_\mu\partial_\mu  \eta^\prime $  differ by a total derivative and hence by a surface term in the effective action. We shall see below that 
it is precisely this surface term that restores the Goldstone boson status of $\eta^\prime$.

It is now obvious that $S=S_0+S_1$ is a very plausible candidate for a low energy effective action satisfying all the basic symmetries of QCD in the chiral limit. But this action results from integrating only the quark degrees of freedom and the gluons still survive in the form of color gauge non-invariant field $K_\mu$. The final effective action must not involve these fields and hence they should be integrated out. The integration of quarks, in reality, would produce a nonlocal action that can be broken into an infinite number of local terms corresponding to various quark loops. But we have considered only those with minimum number of derivatives. Similarly the $K_\mu$ integration would also make the theory non-local and we should filter out the minimal terms.

One way of obtaining it is by contracting the $K_\mu$ fields in the diagrammatic sense \cite{derujula}.

\vspace{12pt}
\begin{figure}[h]
\begin{center}
\includegraphics{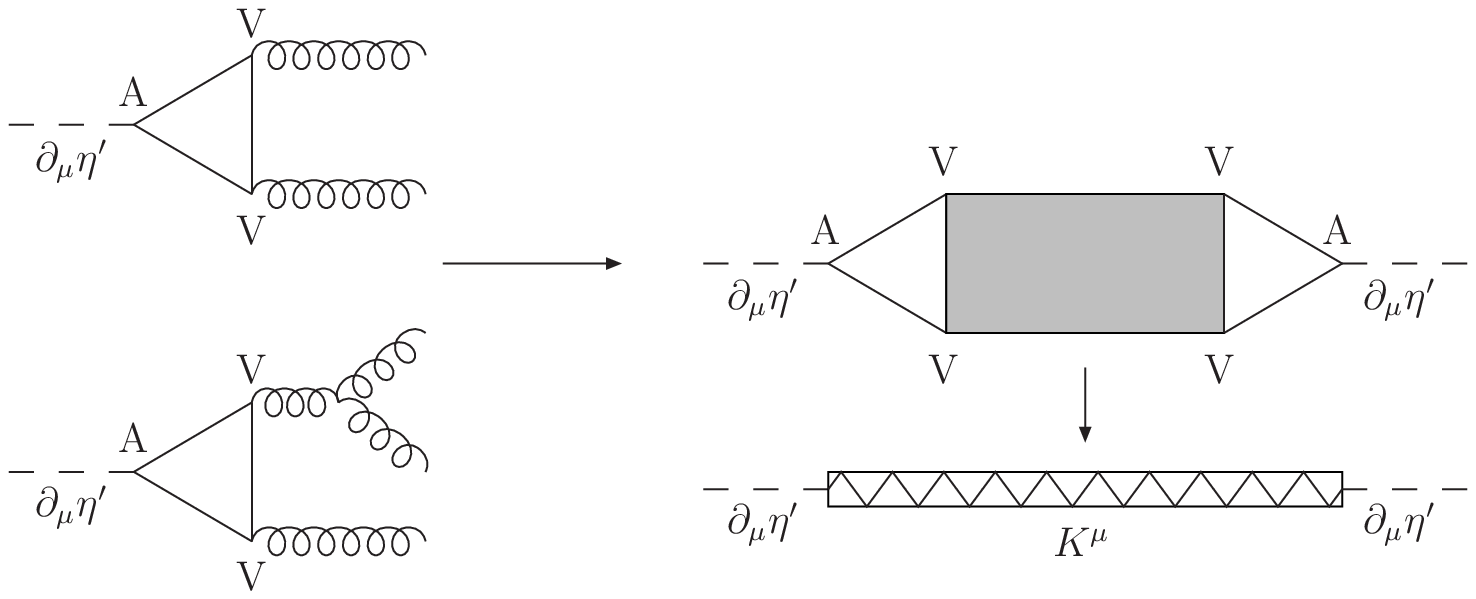}
\end{center}
\end{figure}

In the lowest iteration the contribution of the non-local action would look like
\begin{equation}
S_{NL}=-\frac{L}{2F_\pi^2}\int d^4x\ d^4y\ \partial_\mu\eta^\prime(x) \ D^{\mu\nu}(x-y) \partial_\mu\eta^\prime(y)\label{nonlocal}
\end{equation}
with 
$$
D^{\mu\nu}(x-y)=<0|T(K^\mu(x)K^\nu(y))|0>. 
$$
A priori there is no direct way of evaluating the propagator as we are in the low momentum transfer region of QCD. There are strong indications, however,  that this propagator has a zero mass pole \cite{veneziano1,veneziano,kogutsuskind}. It is connected to the fact that $K_\mu$ does not vanish at infinity since its surface integral is the topological charge.  A zero mass pole would yield a local term  proportional to $(\eta^\prime)^2$ from Eq.(\ref{nonlocal}) \cite{banerjee2} and hence can be identified with a mass term. However, there will be non-local contribution coming from the surface and in the perturbative context is difficult to wirte in any closed form.

A more direct way to integrate out the gluonic components and estimate the surface term is to treat the field $K^\mu$ as a pseudo-gauge field which under color gauge transformation of  gluons $\delta A^a_\mu=f^{abc}\delta\theta^b A_\mu^c +\partial_\mu\delta\theta^a $ transforms as 
\begin{equation}
\begin{array}{ll}
&\delta K_\mu =\displaystyle{\epsilon_{\mu\nu\alpha\beta}\partial^\nu\delta\xi^{\alpha\beta}},\\
{\rm where}\ &\delta\xi^{\alpha\beta}=\displaystyle{A^\alpha_a\partial^\beta\delta\theta_a - \alpha\leftrightarrow\beta \ .}
\end{array}
\end{equation}
The Lagrangian for $K_\mu$ can be written as \cite{aurilia,veneziano}
\begin{equation}
{\cal L}_K=-\frac{L}{\Lambda^4}\left[(\partial_\mu K_\mu)^2+\xi(\partial_\mu K_\nu-\partial_\nu K_\mu)^2\right]
\label{KLagrangian}
\end{equation}
where $\xi$ is the  gauge fixing parameter.
The multiplicative factor  $L/\Lambda^4$ in Eq.(\ref{KLagrangian}), with $\Lambda$ a mass parameter (closely related to the QCD mass scale) has its origin from the susceptibilty caused by  the vacuum polarization of $L$ species of constituent quarks in the Yang-Mills field background.   

The action involving the $\eta^\prime$ and $K_\mu$ fields and their interaction is quadratic and it should be possible to decouple them by a change of variable. To achieve this one has to express the local part of the action in terms of the derivatives of $K_\mu$.  One, therefore, writes the interaction term Eq.(\ref{anomalyterm})  as 
\begin{equation}
{\cal L}_1=\frac{L}{F_\pi}\partial_\mu(\eta^\prime K^\mu)-\frac{L}{F_\pi}\eta^\prime \partial_\mu K^\mu\label{byparts}
\end{equation}
Each term in the right hand side violates chiral symmetry even though the combination remains chiral invariant. The first term, being a total derivative, results in a surface term that does not vanish if $\eta^\prime$ is a massless Goldstone boson. The second term, popular in literature \cite{witten,veneziano,veneziano1} for representing the anomaly  in the effective Lagrangian, breaks chiral symmetry explicitly. The surface term holds the key to the restoration of chiral symmetry.

An essential ingredient for the nontriviality of the surface term is the existence of the zero mode of the $\eta^\prime$ Goldstone boson. It is precisely this zero mode that acts as the generator of $U(1)$ action on the vacuum to make it degenerate \cite{itzykson}. On the surface at infinity only the zero mode (a constant) of $\eta^\prime$ survives and can be brought out of the integral yielding an action which is manifestly nonlocal 
\begin{equation}
\frac{L}{F_\pi}\eta^\prime_0\int d^4x  Q(x)\label{surfaceterm}
\end{equation}

To decouple the modes involving the rest of the action (i.e., the local part of the action of Eq.(\ref{byparts}) along with the kinetic energy terms of $\eta^\prime$ and $K_\mu$) one just needs to
change the functional integration measure from $K_\mu$ to 
\begin{equation}
C_\mu\equiv K_\mu -\partial_\mu\alpha\label{Cmudef},
\end{equation}
with 
\begin{equation}
{\tt \Box}\alpha = \frac{\Lambda^4}{2F_\pi}\eta^\prime.
\end{equation}
This yields the result for the local part of the action 
\begin{equation}
\begin{array}{rl}
S_{\eta^\prime,C_\mu}=&\displaystyle{\int d^4x \left[\frac{1}{2}\partial^\mu\eta^\prime\partial_\mu\eta^\prime -\frac{1}{2}\frac{L\Lambda^4}{2F_\pi^2}(\eta^\prime)^2-\frac{1}{\Lambda^4}\left((\partial_\mu C_\mu)^2+\xi(\partial_\mu C_\nu-\partial_\nu C_\mu)^2\right)\right] }
\label{action2}
\end{array}
\end{equation}
But for the zero mode (Eq.(\ref{surfaceterm})) $\eta^\prime$ is completely delinked from the pseudogauge field $C_\mu$ in the action. The apparent mass
like term in the $\eta^\prime$ action is simply an artifact and the chiral symmetry broken by this is exactly compensated by corresponding chiral symmetry breaking term in the decoupled $C_\mu$ part of the action. 

In reality the pseudoscalar Goldstone particles are not massless. They acquire mass through explicit chiral symmetry breaking in Lagrangian by quark mass 
\begin{equation}
{\cal L}_2=\frac{1}{2}F_\pi^2B Tr({\cal M}^\dagger m + m^\dagger {\cal M})\label{quarkmass}
\end{equation}
where $m$ stands for the light quark mass matrix and $B$ is the condensate mass parameter \cite{leutwyler}.  As soon as the {\it small} quark masses are switched on, the chiral symmetry is explicitly broken and $\eta^\prime$ along with its $L^2-1$ flavour counterparts acquires mass. This would automatically ensure that the pseudoscalar bosons die off at infinity and hence has no zero modes. Thus the nonlocal  part of the action  Eq.(\ref{surfaceterm})decouples from the theory and one is left with the a local action (for the $U(1)$ part)
\begin{equation}
\begin{array}{rl}
S_{\eta^\prime,C_\mu}=&\displaystyle{\int d^4x \left[\frac{1}{2}\partial^\mu\eta^\prime\partial_\mu\eta^\prime -\frac{1}{2}m^2_\pi(\eta^\prime)^2-\frac{1}{2}\frac{L\Lambda^4}{2F_\pi^2}(\eta^\prime)^2-\frac{1}{\Lambda^4}\left((\partial_\mu C_\mu)^2+\xi(\partial_\mu C_\nu-\partial_\nu C_\mu)^2\right)\right] }
\label{action3}
\end{array}
\end{equation}

To summarise, the ansatz for the anomaly term Eq.(\ref{anomalyterm}) reflects and implements the properties that characterise $\eta^\prime$ as the $U(1)$ Goldstone boson in the chiral limit and as
a pseudo-Goldstone boson away from it.
In the chiral limit, global chiral $U(1)$ symmetry guaranteed by Eq.(\ref{conservation}) in the underlying QCD requires invariance under translation  $\eta^\prime\rightarrow\eta^\prime(x)+\theta$. This is realised in the ansatz Eq.(\ref{anomalyterm}), thus ensuring that $\eta^\prime$ remains massless in the chiral limit even in the presence of anomaly. Away from the chiral limit, $\eta^\prime$, along with its flavour counterparts, acquires small mass of order $m_\pi$ arising from the chiral symmetry breaking term Eq.(\ref{quarkmass}) and the translation invariance ceases to be a symmetry in the Goldsone Boson sector. This  is also reflected in the anomaly term written in the form Eq.(\ref{byparts}).  The surface term drops out from the action of $\eta^\prime$. All these properties are incorporated in the action $S_{\eta^\prime,C_\mu}$ of Eq.(\ref{action3}). It also displays explicitly the piece in the $\eta^\prime$ mass that arises exclusively from the anomaly term Eq.(\ref{anomalyterm}). To obtain the total $m_{\eta^\prime}$ one should, of course, add the piece arising from the explicit breaking of the chiral symmetry Eq.(\ref{quarkmass}) that $\eta^\prime$ shares with its flavour counterparts. One thus recovers the Witten-Veneziano formula \cite{witten,veneziano,veneziano1}  for the mass of $\eta^\prime$
\begin{equation}
m_{\eta^\prime}^2=m_\pi^2+ \frac{L\Lambda^4}{2F_\pi^2}\label{massformula}
\end{equation}
Note that the piece induced by anomaly in the $\eta^\prime$ mass, the second term of r.h.s of the Eq.(\ref{massformula}), is triggered only in the presence of the first term arising from explicit breaking of chiral symmetry through quark masses. The anomaly    $A(x)$ cannot and does not    yield, on its own, any mass. Its role is that of a catalytic agent that enhances $\eta^\prime$ mass arising from explicit symmetry breaking terms in the action.  Thus unlike in the large $N_c$ scheme of refs. \cite{witten,veneziano,veneziano1} the singlet pseudo-Golstone boson $\eta^\prime$ that emerges in the present scheme leads to the PCAC relation
\begin{equation}
\partial_\mu J_{\mu 5}=F_\pi m_{\eta^\prime}^2\eta^\prime 
\end{equation}.

\vspace{12pt}

We thank P. Mitra  and A. Harindranath for  discussions.

\end{document}